\documentclass[sigconf]{acmart}

\usepackage{balance}
\usepackage{etoolbox}
\usepackage{color}
\usepackage{footnote}
\usepackage{listings}
\usepackage{tablefootnote}
\usepackage{booktabs}
\usepackage{mathtools}
\usepackage{syntax}
\usepackage{float}
\usepackage{makecell}
\usepackage{array} 
\usepackage{multirow}
\usepackage[title]{appendix}
\usepackage{amssymb}
\usepackage{subfigure}
\usepackage{xcolor}
\usepackage{colortbl}

\makesavenoteenv{tabular}
\makesavenoteenv{table}

\newcolumntype{L}[1]{>{\raggedright\let\newline\\\arraybackslash\hspace{0pt}}m{#1}}
\newcolumntype{C}[1]{>{\centering\let\newline\\\arraybackslash\hspace{0pt}}m{#1}}
\newcolumntype{R}[1]{>{\raggedleft\let\newline\\\arraybackslash\hspace{0pt}}m{#1}}
\newcommand*\rot{\rotatebox{90}}

\newcommand{\TP}{\cellcolor{green!25}\mbox{TP}}
\newcommand{\TN}{\cellcolor{green!25}\mbox{TN}}
\newcommand{\FP}{\cellcolor{red!25}\mbox{FP}}
\newcommand{\FN}{\cellcolor{red!25}\mbox{FN}}

\def\BibTeX{{\rm B\kern-.05em{\sc i\kern-.025em b}\kern-.08emT\kern-.1667em\lower.7ex\hbox{E}\kern-.125emX}}

\newcommand{\toolname}{\textsc{{\AE}GIS}}

\renewcommand\footnotetextcopyrightpermission[1]{} 

\lstdefinelanguage{Rosetta}{
  keywords=[1]{opcode, stack, memory, pc, address, depth, transaction, hash, result, block, number, from},
  keywordstyle=[1]\color{blue},
  keywords=[2]{SSTORE, CALL, DELEGATECALL, CALLCODE, CREATE, CALLDATACOPY, JUMPI, CALLDATALOAD, SELFDESTRUCT, ADD, SUB, MUL, TIMESTAMP, SLOAD},
  keywordstyle=[2]\color{teal},
  keywords=[3]{source, src, destination, dst, where},
  keywordstyle=[3]\color{violet},
}
\usepackage{listings, xcolor}

\definecolor{verylightgray}{rgb}{.97,.97,.97}

\lstdefinelanguage{Solidity}{
	keywords=[1]{anonymous, assembly, assert, balance, break, call, callcode, case, catch, class, constant, continue, contract, debugger, default, delegatecall, delete, do, else, emit, event, export, external, false, finally, for, function, gas, if, implements, import, in, indexed, instanceof, interface, internal, is, length, library, log0, log1, log2, log3, log4, memory, modifier, new, payable, pragma, private, protected, public, pure, push, require, return, returns, revert, selfdestruct, send, storage, struct, suicide, super, switch, then, this, throw, transfer, true, try, typeof, using, value, view, while, with, addmod, ecrecover, keccak256, mulmod, ripemd160, sha256, sha3}, 
	keywordstyle=[1]\color{blue}\bfseries,
	keywords=[2]{address, bool, byte, bytes, bytes1, bytes2, bytes3, bytes4, bytes5, bytes6, bytes7, bytes8, bytes9, bytes10, bytes11, bytes12, bytes13, bytes14, bytes15, bytes16, bytes17, bytes18, bytes19, bytes20, bytes21, bytes22, bytes23, bytes24, bytes25, bytes26, bytes27, bytes28, bytes29, bytes30, bytes31, bytes32, enum, int, int8, int16, int24, int32, int40, int48, int56, int64, int72, int80, int88, int96, int104, int112, int120, int128, int136, int144, int152, int160, int168, int176, int184, int192, int200, int208, int216, int224, int232, int240, int248, int256, mapping, string, uint, uint8, uint16, uint24, uint32, uint40, uint48, uint56, uint64, uint72, uint80, uint88, uint96, uint104, uint112, uint120, uint128, uint136, uint144, uint152, uint160, uint168, uint176, uint184, uint192, uint200, uint208, uint216, uint224, uint232, uint240, uint248, uint256, var, void, ether, finney, szabo, wei, days, hours, minutes, seconds, weeks, years},	
	keywordstyle=[2]\color{teal}\bfseries,
	keywords=[3]{block, blockhash, coinbase, difficulty, gaslimit, number, timestamp, msg, data, gas, sender, sig, value, now, tx, gasprice, origin},	
	keywordstyle=[3]\color{violet}\bfseries,
	identifierstyle=\color{black},
	sensitive=false,
	comment=[l]{//},
	morecomment=[s]{/*}{*/},
	commentstyle=\color{gray}\ttfamily,
	stringstyle=\color{red}\ttfamily,
	morestring=[b]',
	morestring=[b]"
}

\begin{document}

\pagestyle{plain} 

\title{{\AE}GIS: Shielding Vulnerable Smart Contracts Against Attacks}

\author{Christof Ferreira Torres}
\affiliation{%
  \institution{SnT, University of Luxembourg}
  \city{Luxembourg}
  \country{Luxembourg}}
\email{christof.torres@uni.lu}

\author{Mathis Baden}
\affiliation{%
  \institution{SnT, University of Luxembourg}
  \city{Luxembourg}
  \country{Luxembourg}}
\email{mathis.steichen@uni.lu}

\author{Robert Norvill}
\affiliation{%
  \institution{SnT, University of Luxembourg}
  \city{Luxembourg}
  \country{Luxembourg}}
\email{robert.norvill@uni.lu}

\author{Beltran Borja Fiz Pontiveros}
\affiliation{%
  \institution{SnT, University of Luxembourg}
  \city{Luxembourg}
  \country{Luxembourg}}
\email{beltran.fiz@uni.lu}

\author{Hugo Jonker}
\affiliation{%
  \institution{$^1$Open University of the Netherlands}
  \city{Heerlen}
  \country{Netherlands}
}
\affiliation{%
  \institution{$^2$Radboud University}
  \city{Nijmegen}
  \country{Netherlands}
}
\email{hugo.jonker@ou.nl}

\author{Sjouke Mauw}
\affiliation{%
  \institution{SnT, University of Luxembourg}
  \city{Luxembourg}
  \country{Luxembourg}}
\email{sjouke.mauw@uni.lu}

%

%
\begin{abstract}
In recent years, smart contracts have suffered major exploits, costing
millions of dollars. Unlike traditional programs, smart contracts are
deployed on a blockchain. 
As such, they cannot be modified once deployed.
Though various tools have been proposed to detect vulnerable smart
contracts, the majority fails to protect vulnerable contracts that have already been deployed on the blockchain.
Only very few solutions have been proposed so far to tackle the issue of post-deployment.
However, these solutions suffer from low precision and are not generic enough to prevent any type of attack.

In this work, we introduce \toolname{}, a dynamic analysis tool
that protects smart contracts from being
exploited during runtime. 
Its capability of detecting new vulnerabilities can easily be extended through so-called attack patterns. 
These patterns are written in a domain-specific language that is
tailored to the execution model of Ethereum smart contracts. The language enables
the description of malicious control and data flows.
In addition, we propose a novel mechanism to streamline and speed up the process of managing attack patterns. 
Patterns are voted upon and stored via a smart contract, thus
leveraging the benefits of tamper-resistance and transparency provided
by the blockchain.
We compare \toolname{} to current state-of-the-art tools and demonstrate that our solution achieves higher precision in detecting attacks.
Finally, we perform a large-scale analysis on the first 4.5 million blocks of the Ethereum blockchain, thereby confirming the occurrences of well reported and yet unreported attacks in the wild.
\end{abstract}

%
\keywords{Ethereum; Smart contracts; Exploit prevention; Security updates}

%
\maketitle

\section{Introduction}

Blockchain has evolved greatly since its first introduction in 2009~\cite{nakamoto2008bitcoin}.
A blockchain is essentially a verifiable, append-only list of records in
which all transactions are recorded in batches of so-called blocks. Each
block is linked to a previous block via a cryptographic hash. This linked list
of blocks is maintained by a decentralised peer-to-peer network. The peers
in this network follow a consensus protocol that dictates which peer is allowed
to append the next block. By introducing the concept of smart contracts,
Ethereum~\cite{wood2014ethereum} revolutionized the way digital assets are
traded. As smart contracts govern more and more valuable assets, the contracts
themselves have come under attack from hackers.

Smart contracts are programs that are stored and executed across blockchain
peers. They are deployed and invoked via transactions. Deployed smart
contracts are immutable, thus any bugs present during
deployment~\cite{atzei2017}, or as a result of changes to the blockchain
protocol~\cite{chainsecurity2019reentrancy}, can make a smart contract
vulnerable. Moreover, since contract owners are anonymous, responsible
disclosure is usually infeasible or very hard in practice. Though smart
contracts can be implemented with upgradeability and destroyability in
mind, this is not compulsory. 
As a matter of fact, Ethereum already faced several devastating attacks
on vulnerable smart contracts.

In 2016, an attacker exploited a reentrancy bug in a
crowdfunding smart contract known as the DAO. The attacker exploited the
capability of recursively calling a payout function contained in the
contract.
The attacker managed to drain over \$150
million~\cite{siegel2016understanding} worth of cryptocurrency from the
smart contract.
The DAO hack was a poignant demonstration of
the impact that insecure smart contracts can have. The Ethereum market
cap value dropped from over \$1.6 billion before the attack, to values
below \$1 billion after the attack, in less than a day. Another
example happened with the planned Constantinople hard fork in January
2019. Ethereum was scheduled to receive an update intended to introduce a
cheaper gas cost for certain smart contract operations. On the eve of
the hard fork, a new reentrancy issue caused by this update was detected.
It turned out that the reduction of gas costs also enabled reentrancy
attacks on smart contracts that were previously secure. This resulted
in the update being delayed~\cite{chainsecurity2019reentrancy}.
A third example is the Parity wallet hack. In 2017, the Parity wallet
smart contract was attacked twice due to a bug in the access control
logic. The bug allowed anyone to claim ownership of the smart
contract and to take control of all the funds. The first attack resulted
in over \$30 million being stolen~\cite{zhao2017parity}, whereas the
second attack resulted in roughly \$155 million being locked
forever~\cite{petrov2017another}.

The manner in which these issues are currently handled is not ideal. At the moment, whenever a major vulnerability is detected by the Ethereum community, it can take several days or weeks for the community to issue a critical update and even longer for all nodes to adopt this update. Such a delay extends the
window for exploitation and can have dire effects on the trading value
of the underlying cryptocurrency. 
Moreover, the lack of a standardised procedure to deal with vulnerable smart contracts, has led to a ``Wild West''-like situation where several self-appointed white hats started attacking smart contracts in order to protect the funds that are at risk from other malicious attackers~\cite{whitehat}.
However, in some cases the effects of attacks can cause a split in the community so contentious that it leads to a hard fork,
such as in the case of the DAO hack which led to
the birth of the Ethereum classic blockchain~\cite{siegel2016understanding}.

Academia has proposed a plethora of different tools that allow users to
scan smart contracts for vulnerabilities prior to deploying them on the
blockchain or interacting with them (see
e.g.~\cite{Luu2016,krupp2018teether,torres2018osiris,tsankov2018securify}).
However, by design these tools cannot protect
vulnerable contracts that have already been deployed.
Grossman et al.~\cite{grossman2017online} are the first to present \textsc{ECFChecker}, a tool that allows to dynamically check executed transactions for reentrancy. 
However, \textsc{ECFChecker} does not prevent reentrancy attacks. 
In order to protect already deployed contracts, Rodler et
al.~\cite{rodler2019sereum} propose \textsc{Sereum}, a modified Ethereum
client that detects and reverts\footnote{Consuming gas,
without letting the transaction affect the state of the blockchain.} transactions that trigger reentrancy
attacks. \textsc{Sereum} leverages the principle that every exploit is
performed via a transaction. Unfortunately, \textsc{Sereum} has three
major drawbacks. First, it requires the client to be modified whenever a
new type of vulnerability is found. Second,
not only the tool itself, but also any updates to it
must be manually adopted by the majority of nodes for its security
provisions to become effective. Third, their detection technique can
only detect reentrancy attacks, despite there being many other types of
attacks~\cite{atzei2017}.

\noindent
\newline
In summary, we make the following contributions:
\begin{itemize}
\item We introduce a novel domain-specific language, which enables
    the description of so-called \emph{attack patterns}. These patterns
    reflect malicious control and data flows that occur during execution
    of malicious transactions.
\item We present \toolname{}, a tool that reverts malicious transactions
    in real-time using attack patterns, thereby preventing attacks on
    deployed smart contracts.
\item We propose a novel way to quickly propagate security updates
    without relying on client-side update mechanisms, by making
    use of a smart contract to store and vote upon new attack patterns.
    Storing patterns in a smart contract ensures integrity, decentralizes
    security updates and provides full transparency on the proposed patterns. 
\item We illustrate the effectiveness by providing patterns to prevent
    the two most prominent hacks in Ethereum, the DAO and Parity wallet hacks. 
\item Finally, we provide a detailed comparison to current state-of-the-art
    runtime detection tools and perform a large-scale analysis on 4.5 million
    blocks. The results demonstrate that \toolname{} achieves better precision
    than current state-of-the-art tools.
\end{itemize}

\section{Background}

In this section, we provide the necessary background for understanding the setting of our work. 
We describe the Ethereum block-chain and its capability of executing smart contracts.
We focus on Ethereum since it is currently the most prominent blockchain platform when it comes to smart contract deployment.
Finally, we also provide background information on the two most prominent smart contract vulnerabilities, namely, reentrancy and access control.

\subsection{Ethereum and Smart Contracts}
\textbf{Ethereum.}
The Ethereum blockchain is a decentralized public ledger that is maintained by a network of nodes that distrust one another.
Every node runs one of several existing Ethereum clients.
The clients can operate with different configurations.
For instance, nodes who are configured to mine blocks are called miners.
Miners execute transactions, include them in blocks and append them to the blockchain.
They compete to create a block by solving a cryptographic puzzle.
Once they succeed, the block is proposed to the network.
Other miners verify the new block and either accept or reject it.
A miner whose block is included in the blockchain is rewarded with a block reward and the execution fees from the included transactions.

\noindent
\newline
\textbf{Transactions.}
Transactions are used to modify state in Ethereum.
As such, they allow users to transfer ether (Ethereum's cryptocurrency), and to create smart contracts or trigger their execution.
Transactions are created using an account.
There are two types of accounts in Ethereum, user accounts and contract accounts.
Transactions are given a certain amount of gas to execute, called the gas limit.
Gas is a unit which is used to measure the use of computing resources.
Gas can be converted to ether through the so-called gas price of a transaction.
Gas limit and gas price can be chosen by the creator of the transaction.
Together they determine the fee that the user is willing to pay for the inclusion of their transaction into the blockchain.
Moreover, transactions also contain a destination address.
It identifies the recipient of the transaction, and it can be either a user account or a smart contract.
Transactions can also carry value that is transferred to the recipient.
Once created, transactions are broadcast to the network.
Miners then execute the transactions and include them into blocks.
Smart contracts (i.e. contract accounts) 
are created by leaving the destination address of a transaction empty.
The bytecode that is provided within the transaction is then copied into the blockchain and it is given a unique address that identifies the smart contract.

\noindent
\newline
\textbf{Smart Contracts.}
Smart contracts are fully-fledged programs that are stored and executed across the blockchain.
They are developed using a dedicated high-level programming language that compiles into low-level bytecode. This bytecode gets interpreted by the Ethereum Virtual Machine.
Smart contracts contain functions that can be triggered via transactions.
The name of the function as well as the data to be executed is included in the data field of the transaction.
A default function or so-called fallback function is executed whenever the provided function name is not recognized by the smart contract.
Moreover, smart contracts can initiate calls to other smart contracts.
Thus, a single transaction may interact with several smart contracts that call one another.
By default smart contracts cannot be destroyed or updated.
It is the task of the developer to implement these capabilities before deploying the smart contract.
Unfortunately, many smart contracts are released without destroyability or upgradeability in mind.
As a result, many contracts remain vulnerable or active on the blockchain even past their utility.
As mentioned earlier, once deployed, smart contracts are immutable, they cannot be modified and bugs cannot be fixed.
Thus, it is not possible to update a smart contract in the later run.

\noindent
\newline
\textbf{EVM.}
The Ethereum Virtual Machine (EVM) is a purely stack-based, register-less virtual machine that supports a Turing-complete instruction set of opcodes.
These opcodes allow smart contracts to perform memory operations and interact with the blockchain, such as retrieving specific information (e.g., the current block number).
Ethereum makes use of gas to make sure that contracts terminate and to prevent denial-of-service attacks.
It assigns a gas cost to the execution of an opcode.
The execution of a smart contract results in the modification of its state.
The latter is stored on the blockchain and consists of a balance and a storage.
The balance represents the amount of ether currently owned by the smart contract.
The storage is organized as a key-value store and allows the smart contract to store values and keep state across executions.
During execution, the EVM holds a machine state $\mu = (g, pc, m, i, s)$, where $g$ is the gas available, $pc$ is the program counter, $m$ represents the memory contents, $i$ is the active number of words in memory and $s$ is the content of the stack.
In summary, the EVM is a transaction-based state machine that updates a smart contract based on transaction input data and the smart contract's bytecode.

\subsection{Smart Contract Vulnerabilities}
\label{sec:reentrancy}
Although, a number of smart contract vulnerabilities exist~\cite{atzei2017}, in this work, we primarily focus on two types of vulnerabilities that have been defined by the NCC Group as the top two vulnerabilities in their Decentralized Application Security Project~\cite{dasp}: \textit{reentrancy} and \textit{access control}.

\begin{figure}
\begin{centering}
\begin{lstlisting}[language=Solidity]
contract A { // Victim contract
  ...
  function withdraw() public {
    if (credit[msg.sender]) {
      msg.sender.call.value(credit[msg.sender])();
      credit[msg.sender] = 0;
    }
}

contract B { // Exploiting contract
  ...
  function () public payable {
    A.withdraw();
  }
}
\end{lstlisting}
\caption{Example of a reentrancy vulnerability. 
}
\label{fig:reentrancy}
\end{centering}
\end{figure}

\noindent
\newline
\textbf{Reentrancy Vulnerabilities.}
Reentrancy occurs whenever a contract calls another contract, which then calls back into the original contract,
thereby creating a reentrant call. This is not an issue as long as all
the state updates that depend on the call from the original contract
are performed before the call. In other words, reentrancy only becomes
problematic when a contract updates its state after calling another
contract. A malicious contract can take advantage of this by recursively
calling a contract until all the funds are drained.
Figure~\ref{fig:reentrancy} provides an example of a malicious
reentrancy. Contract $B$ contains a fallback function (line 12-14), a
default function that is automatically executed when no other function
is called. In this example, the fallback function of contract $B$ calls
the withdraw function of contract $A$. Assuming that contract $B$
already deposited some ether in contract $A$, contract $A$ now calls
contract $B$ to transfer back its deposited ether. However, the transfer
results in calling the fallback function of contract $B$ once again,
which results in reentering contract $A$ and once more transferring the
value of the deposited ether to contract $B$. This repeats until
the balance of contract $A$ becomes zero or the execution
runs out of gas. 

Reentrancy vulnerabilities were extensively studied by Rodler et
al.~\cite{rodler2019sereum}, and can be divided into four distinct
categories: \emph{same-function} reentrancy, \emph{cross-function}
reentrancy, \emph{delegated} reentrancy and \emph{create-based}
reentrancy. Same-function reentrancy occurs whenever an attacker
reenters the original contract via the same function (see
Figure~\ref{fig:reentrancy}). Cross-function reentrancy builds on the
same-function reentrancy. However, here the attacker takes advantage of
another function that shares a state with the original function.
Delegated reentrancy and create-based reentrancy are similar to
same-function reentrancy, but use different opcodes to initiate the
call. Specifically, delegated reentrancy can occur using either the
\texttt{DELEGATECALL} or \texttt{CALLCODE} opcodes, while create-based
reentrancy only occurs when using the \texttt{CREATE} opcode. While the
\texttt{DELEGATECALL} and \texttt{CALLCODE} opcodes behave roughly
similar to the \texttt{CALL} opcode, the \texttt{CREATE} opcode causes a new
contract to be created and the contract constructor to be executed. This
newly created contract can then call and reenter the original contract.

\begin{figure}[t]
\begin{centering}
\begin{lstlisting}[language=Solidity]
contract W { // Wallet contract 
  ...
  function W(address _owner) { // Contructor
    L.delegatecall("initWallet(address)", _owner);
  }
  function () payable {
    L.delegatecall(msg.data);
  }
}

contract L { // Library contract
  ...
  modifier onlyOwner {
    if (m_ownerIndex[msg.sender] > 0) _;
  }
  ...
  function initWallet(address[] _owners, uint _required, uint _daylimit) {
    initDaylimit(_daylimit);
    initMultiowned(_owners, _required);
  }
  function initMultiowned(address[] _owners, uint _required) {
    ...
    for (uint i = 0; i < _owners.length; ++i) {
      ...
      m_ownerIndex[_owners[i]] = 2+i;
    }
    ...
  }
  function execute(address _to, uint _value, bytes _data) onlyOwner {
    _to.call.value(_value)(_data));
  }
  function kill(address _to) onlyOwner {
    suicide(_to);
  }
}
\end{lstlisting}
\caption{Example of an access control vulnerability.}
\label{fig:acl}
\end{centering}
\end{figure}

\noindent
\newline
\textbf{Access Control Vulnerabilities.}
Access control vulnerabilities result from incorrectly enforced user access control policies in smart contracts.
Such vulnerabilities allow attackers to gain access to privileged contract functions that would normally not be available to them.
The most famous examples of this type of vulnerability are the two Parity MultiSig-Wallet hacks~\cite{zhao2017parity,petrov2017another}. 
The issue originates from the fact that the developers of the Parity wallet decided to split some of the contract logic into a separate smart contract named \texttt{WalletLibrary}. This had the advantage of reusing parts of the code for multiple wallets allowing users to save on gas costs during deployment.
A simplified version of the code can be seen in Figure~\ref{fig:acl}. As can be seen in line 17-20, the initialisation of the wallet is performed via the \texttt{initWallet} function located in contract $L$, which is called by the constructor of contract $W$. In addition, any unmatched function calls to contract $W$ are caught by the fallback function in line 6-8, which redirects the call to contract $L$ by means of the \texttt{DELEGATECALL} operation.
Unfortunately, in the first version of the Parity MultiSig-Wallet, the developers forgot to write a safety check for the \texttt{initWallet} function, ensuring that the function can only be called once. 
As a result an attacker was able to gain ownership of contract $W$ by calling the \texttt{initWallet} function via the fallback function. Once in control the attacker withdrew all the funds by invoking the \texttt{execute} function (line 32-34).

After the first Parity hack, a new Parity MultiSig-Wallet Library contract was deployed addressing the issue above.
In the newly deployed version, the \texttt{initWallet} function was not part of the constructor anymore, but had to be called separately after deployment.
However, the developers did not call the \texttt{initWallet} function after deployment.
Hence, contract $L$ remained uninitialised, meaning that the library contract itself had no owners.
As a result, 3 months after deployment a user known as \emph{devops199} was experimenting with the previous Parity hack vulnerability and called the \texttt{initWallet} function directly inside contract $L$, marking its address as the owner. The user then called the \texttt{kill} function (line 32-34), which removed the executable code of contract $L$ from the blockchain\footnote{The contract code is technically not removed from the blockchain, however, the contract's code can no longer be executed on the blockchain, because the contract has been marked as killed.} and sent the remaining funds to the new owner. The contract itself contained no funds, however it was used by multiple Parity wallets which had the address of contract $L$ defined as a constant in their executable code. As a result any wallet trying to use contract $L$ as a library would now receive zero as return value, effectively rendering the wallet unusable and therefore freezing the funds contained in the wallets.
This led the user to publicly disclose the steps that led to this tragedy, with the words: ``I accidentally killed it.''~\cite{devops199}.

\section{Related Work}

In this section, we discuss some of the works that are most closely related to ours.

\noindent
\newline
\textbf{Security Analysis of Smart Contracts.}
As with any program, smart contracts may contain bugs and can be vulnerable to exploitation.
As discussed in \cite{atzei2017}, different types of vulnerabilities exist, often leading to financial losses.
The issue is made worse by the fact that smart contracts are immutable.
Once deployed, they cannot be altered and vulnerabilities cannot be fixed.
In addition to that, automated tools for launching attacks exist~\cite{krupp2018teether}.

Several defense mechanisms have been proposed to detect security vulnerabilities in smart contracts. 
This includes tools such as \textsc{Erays} \cite{zhou2018erays}, designed to provide smart contract auditors with a reverse engineered pseudo code of a contract from its bytecode.
The interpretation of the pseudo code however remains a slow and gruelling task.
More automated tools have also been proposed benefiting from regular expressions \cite{zhang2019soliditycheck} and machine learning techniques \cite{tann2018towards} to detect vulnerabilities.

A wealth of security research has focused on the creation of static analysis tools to automatically detect vulnerabilities in smart contracts.
Formal verification has been used together with a formal definition of the EVM \cite{hildenbrandt2018kevm,amani2018towards}, or by first converting smart contracts into the formal language F* \cite{bhargavan2016formal, grishenko2018a}.
Other works focused on analysing the higher level solidity code \cite{tikhomirov2018smartcheck,feist2019slither}, which limits the scope to those contracts with available source code.
Another approach is to apply static analysis on the smart contract bytecode \cite{tsankov2018securify}.
A technique commonly used for this purpose is symbolic execution, designed to thoroughly explore the state space of a smart contract utilising constraint solving.
It has been used to detect contracts with vulnerabilities \cite{Luu2016,permenev2020verx}, to find misbehaving contracts \cite{nikolic2018finding,kolluri2019exploiting,torres2019honeybadger}, or 
detect integer bugs \cite{torres2018osiris,kalra2018zeus}.
Fuzzing techniques have also been applied \cite{jiang2018contractfuzzer,he2019learning}.
In \cite{wustholz2019harvey} the authors propose \textsc{Harvey}, a greybox fuzzer that selects appropriate inputs and transaction sequences to increase code coverage.
Fuzzing techniques however involve a trade-off between the number of discovered paths and the efficiency in input generation.

While all the listed tools help identify vulnerabilities, they cannot protect already deployed smart contracts from being exploited. Therefore, to deal with the issue of vulnerabilities in deployed smart contracts, \cite{grossman2017online,rodler2019sereum} propose a modification to the Ethereum client, that would allow detection and prevent exploitation of reentrancy vulnerabilities at runtime.
However, these approaches only deal with reentrancy and require all the clients in the network to be modified.
This is an issue for the following reasons.
On one hand, every update of the vulnerability detection software requires an update of the different Ethereum client implementations.
This is true for both bug fixes and functionality upgrades, for example the detection of new vulnerabilities.
On the other hand, every modification of the clients needs to be adopted by all the nodes participating in the Ethereum blockchain.
This requires 
time and breaks compatibility between updated and non-updated clients.
In this work, we propose a generic solution that 
protects contracts and users from existing and future vulnerabilities, without requiring client modifications and forks every time a new vulnerable smart contract is found.

Wang et al.~\cite{wang2019vultron} propose an approach to detect vulnerabilities at runtime based on two invariants that follow the intuition that most vulnerabilities are due to a mismatch between the transferred amount and the amount reflected by the contract's internal bookkeeping logic. 
However, this approach has three main drawbacks. 
First, it requires the automated and correct identification of bookkeeping variables, which besides being a non-trivial task also does not hold for every contract, since there can be contracts that do not use internal bookkeeping logic but are nevertheless vulnerable.
Second, their approach does not model environmental information such as timestamps or block numbers, which does not allow them to detect vulnerabilities such as timestamp dependence or transaction order dependency, whereas our approach models environmental information and allows for the detection of these vulnerabilities.
Finally, Wang et al.'s approach can only detect violations of safety properties and not violations of liveness properties such as the Parity Wallet Hack 2. In this work, we demonstrate that our approach is capable of detecting both Parity wallet hacks and therefore violations to safety as well as liveness properties.

\noindent
\newline
\textbf{Blockchain-Based Voting.}
Since blockchains provide the means for transparency and decentralization, multiple blockchain-based solutions have been proposed for performing electronic voting~\cite{osgood2016future,ayed2017conceptual,hjalmarsson2018blockchain}. Interestingly, with the recent developments in quantum computers, recent work also has started to focus on the development of quantum-resistant blockchain-based voting schemes \cite{sun2019simple}.
These solutions can all be categorised into two categories: cryptocurrency-based and smart-contract-based.

Cryptocurrency-based solutions focus on using payments as a proxy for votes in an election. When a voter wishes to cast a vote, he or she makes a payment to the address of the candidate. Lee et al.~\cite{lee2016electronic} proposed such a system in the Bitcoin network. However, their system requires a trusted third party to perform the ballot counting. Zao et al.~\cite{zhao2015vote} were the first to propose a voting scheme using the public Bitcoin network while preserving the privacy of the votes. Another well-known cryptocurrency-based solution is CarbonVote \cite{carbonvote}. It was introduced in the aftermath of the DAO hack to allow the Ethereum Foundation to determine if the Ethereum community wanted a hard fork or not. The tallying was performed by counting the amount of ether that each address received. Needless to say, such a system gives a tremendous amount of voting power to users with a large amount of funds.

Smart-contract-based voting relies on a decentralized application to assist the voting process -- there is no central entity. McCorry et al.~\cite{mccorry2017smart} propose a practical implementation of the Open Vote Network~\cite{hao2010anonymous} in the form of a smart contract deployed on the Ethereum blockchain for boardroom voting.
Their implementation is self-tallying and provides, in addition to vote privacy, also transparency. Voting proceeds in several rounds, where the voters first broadcast their voting key, followed by a proof that their vote is binary (a ``yes'' or ``no'' vote).  A final tally round allows anyone to calculate the total sum of votes, without revealing individual ballots.
The voting mechanism described in this paper is inspired by McCorry et al.'s proposed solution and implementation. The limitations of their proposed solution, namely having a binary voting system and limiting the number of voters to less than 50 participants, are acceptable for our purposes.

\section{Methodology}

In this section, we present the details of our solution towards a generic and decentralized way to prevent any type of attacks on already deployed smart contracts.
Our idea is to bundle every Ethereum client with a runtime analysis tool, that interacts with the EVM and is capable of interpreting so-called \textit{attack patterns}, and reverting transactions that match these patterns.
Attack patterns are described using our domain-specific language (DSL), which is tailored to the execution model of the EVM and which allows to easily describe malicious control and data flows.
The fact that we shift the capability of detecting attacks from the client-side implementation to the DSL, gives us the advantage of being able to quickly propose mitigations against new vulnerabilities, without having to modify the Ethereum client.
Existing approaches, such as \textsc{Sereum} for example, require the client-side implementation to be modified whenever a new vulnerability is found.

\subsection{Generic Attack Detection}

Attacks are detected in our system through the use of patterns, which are described using our DSL.
The DSL allows for the definition of malicious events that occur during the execution of EVM instructions.
The syntax of our DSL is defined by the following BNF grammar:

\setlength{\grammarindent}{6.0em} 

\begin{figure}[H]
\centering
\begin{grammar}
<instr> ::= \texttt{CALL} | \texttt{CALLDATALOAD} | \texttt{SSTORE} | \texttt{JUMPI} | $\dots$

<exec> ::= \textbf{depth} | \textbf{pc} | \textbf{address} | \textbf{stack(}int\textbf{)} | \textbf{stack.result} | 
\alt \textbf{memory(}int, int\textbf{)} | \textbf{transaction.}<trans> 
\alt \textbf{block.}<block>

<trans> ::= \textbf{hash} | \textbf{value} | \textbf{from} | \textbf{to} | $\dots$

<block> ::= \textbf{number} | \textbf{gasUsed} | \textbf{gasLimit} | $\dots$

<comp> ::= \textless\thinspace | \textgreater\thinspace | $\leq$ | $\geq$ | $=$ | $\neq$ | + | - | $\cdot$ | /

<expr> ::= \textbf{(src.}<exec> <comp> <expr>\textbf{)} [$\wedge$ <expr>]
\alt \textbf{(}<expr> <comp> \textbf{dst.}<exec>\textbf{)} [$\wedge$ <expr>]
\alt \textbf{(src.}<exec> <comp> \textbf{src.}<exec>\textbf{)} [$\wedge$ <expr>]
\alt \textbf{(src.}<exec> <comp> \textbf{dst.}<exec>\textbf{)} [$\wedge$ <expr>]
\alt \textbf{(dst.}<exec> <comp> \textbf{dst.}<exec>\textbf{)} [$\wedge$ <expr>]
\alt \textbf{(src.}<exec> <comp> int\textbf{)} | \textbf{(dst.}<exec> <comp> int\textbf{)} 

<rel> ::= $\Rightarrow$ | $\leadsto$ | $\rightarrow$

<pattern> ::= \textbf{(opcode} = <instr>\textbf{)} <rel> \textbf{(opcode} = <instr>\textbf{)} [\textbf{where} <expr>]
\alt <pattern> <rel> \textbf{(opcode} = <instr>\textbf{)} [\textbf{where} <expr>]
\alt \textbf{(opcode} = <instr>\textbf{)} <rel> <pattern> [\textbf{where} <expr>]
\end{grammar}
\caption{DSL for describing attack patterns.}
\label{fig:syntax}
\end{figure}

\begin{figure*}
  \centering
  \includegraphics[scale=0.52]{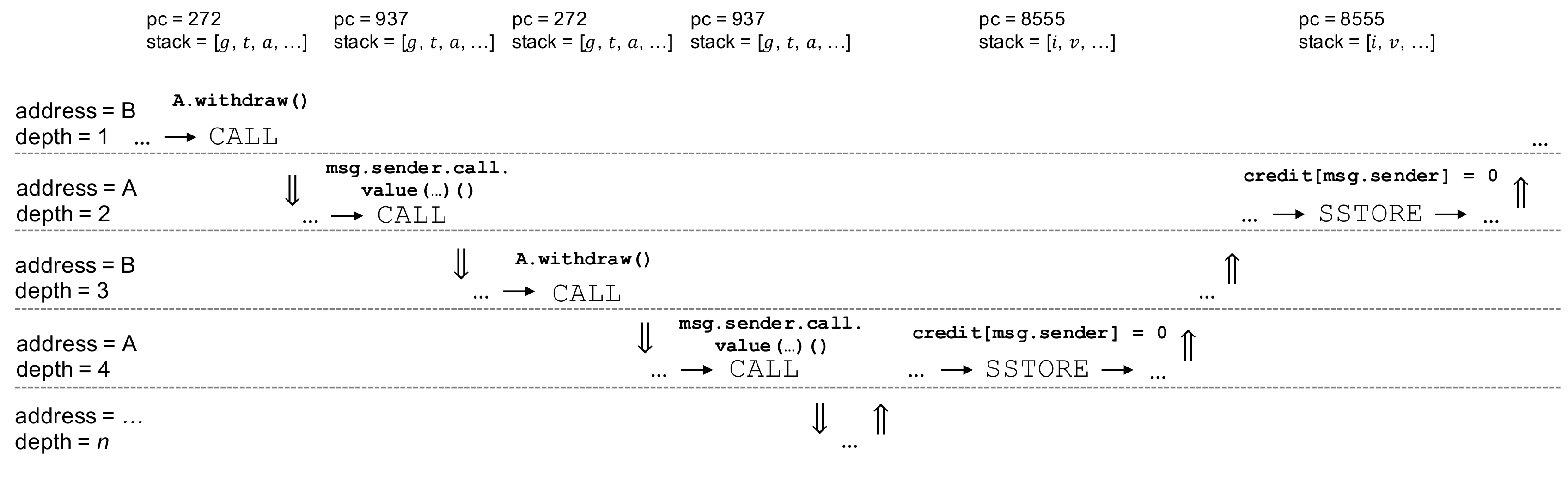}
  \caption{Execution example of a reentrancy attack, where the stack values $g$ (gas), $t$ (to), $a$ (amount), $i$ (index) and $v$ (value) represent the respective parameters passed to the instructions during execution. A control flow relation is depicted using $\Rightarrow$, while $\rightarrow$ depicts a follows relation.}
  \label{fig:instructionflow}
\end{figure*}

A pattern is a sequence of relations between EVM instructions that may occur at runtime. 
We distinguish between three types of relations, a ``control flow'' relation ($\Rightarrow$), a ``data flow'' relation ($\leadsto$), and a ``follows'' relation ($\rightarrow$). 
A control flow relation means that an instruction is control dependent on another instruction.
A data-flow relation means that an instruction is data dependent on another instruction.
A  follows relation means that an instruction is executed after another instruction, without necessarily being control or data dependent on the other instruction. 
A relation is always between two EVM opcodes: a source opcode (\textcolor{violet}{\texttt{src}}) and a destination opcode (\textcolor{violet}{\texttt{dst}}). The source marks the beginning of the relation, whereas the destination defines the end of the relation.
Moreover, the DSL allows to create conjunctions of expressions that allow to compare the execution environment between source and destination.
The execution environment includes the current depth of the call stack (\textcolor{blue}{\texttt{depth}}), the current value of the program counter (\textcolor{blue}{\texttt{pc}}), the address of the contract that is currently being executed (\textcolor{blue}{\texttt{address}}), the current values on the stack (\textcolor{blue}{\texttt{stack}}) as well as the result of an operation that is pushed onto the stack (\textcolor{blue}{\texttt{stack.result}}), the current values stored in memory (\textcolor{blue}{\texttt{memory}}), and finally, properties of the current transaction that is being executed (e.g. \textcolor{blue}{\texttt{hash}}) as well as properties of the current block that is being executed (e.g. \textcolor{blue}{\texttt{number}}).
The stack is addressable via an integer, where 0 defines the top element on the stack.
The memory is addressable via two integers: an offset and a size.
In the following, we explain the semantics of our DSL via two concrete examples of attack patterns: \emph{same-function reentrancy} and the \emph{parity wallet hack 1}.

\noindent
\newline
\textbf{Same-Function Reentrancy.}
Reconsider the reentrancy example that was described in Section~\ref{sec:reentrancy}.
Figure~\ref{fig:instructionflow}, illustrates the control flow as well as the follows relations that occur during the execution of that example.
The execution starts with contract address $B$ and a call stack depth of 1.
Eventually, contract $B$ calls the withdraw function of contract $A$, which results in executing the \textcolor{teal}{\texttt{CALL}} instruction and increasing the depth of the call stack to 2, and switching the address of the contract that is being executed to contract $A$.
Next, contract $A$ sends some funds to contract $B$, which also results in executing the \textcolor{teal}{\texttt{CALL}} instruction and increasing the depth of the call stack to 3, and switching the address of the contract that is being executed back to contract $B$. 
As a result, the fallback function of contract $B$ is called, which in turn calls again the withdraw function of contract $A$.
This sequence of calls repeats until the balance of contract $A$ is either empty or the execution runs out of gas.
Eventually, the state in contract $A$ is updated by executing the \textcolor{teal}{\texttt{SSTORE}} instruction.
Given these observations, we can now create the following attack pattern in order to detect and thereby prevent same-function reentrancy:

\lstnewenvironment{rosettatable}{
    \lstset{language=rosetta,
    basicstyle=\footnotesize,
    numbers=none,
    xleftmargin=0pt,
    xrightmargin=0pt
    }%
  }
  {}

\begin{rosettatable}
(opcode = CALL) $\Rightarrow$ (opcode = CALL) where 
 (src.stack(1) = dst.stack(1)) $\wedge$ 
 (src.address = dst.address) $\wedge$
 (src.pc = dst.pc) $\rightarrow$ 
(opcode = SSTORE) $\rightarrow$ (opcode = SSTORE) where 
 (src.stack(0) = dst.stack(0)) $\wedge$ 
 (src.address = dst.address) $\wedge$ 
 (src.depth > dst.depth)
\end{rosettatable} 

\noindent
\newline
This attack pattern evaluates to true if a transaction meets the following two conditions: 
\begin{enumerate}
\item there is a control flow relation between two \textcolor{teal}{\texttt{CALL}} instructions, where both instructions share the same call destination (i.e.~\texttt{\textcolor{violet}{src}.
\textcolor{blue}{stack}(1) = \textcolor{violet}{dst}.\textcolor{blue}{stack}(1)}), are executed by the same contract (i.e.~\texttt{\textcolor{violet}{src}.\textcolor{blue}{address} = \textcolor{violet}{dst}.\textcolor{blue}{address}}) and share the same program counter (i.e.~\texttt{\textcolor{violet}{src}.\textcolor{blue}{pc} = \textcolor{violet}{dst}.\textcolor{blue}{pc}});
\item two \textcolor{teal}{\texttt{SSTORE}} instructions follow the previous control flow relation, where both instructions write to the same storage location (i.e.~\texttt{\textcolor{violet}{src}.\textcolor{blue}{stack}(0) = \textcolor{violet}{dst}.\textcolor{blue}{stack}(0)}), are executed by the same contract (i.e.~\texttt{\textcolor{violet}{src}.\textcolor{blue}{address} = \textcolor{violet}{dst}.\textcolor{blue}{address}}) and where the first instruction has a higher call stack depth than the second instruction (i.e.~\texttt{\textcolor{violet}{src}.\textcolor{blue}{depth} $>$ \textcolor{violet}{dst}.\textcolor{blue}{depth}}).
\end{enumerate}

\noindent
It is worth mentioning that we compare the program counter values of the two \textcolor{teal}{\texttt{CALL}} instructions in order to make sure that it is the same function that is being called, as our goal is to detect only same-function reentrancy.

\noindent
\newline
\textbf{Parity Wallet Hack 1.}
\begin{figure*}
  \centering
  \includegraphics[scale=0.47]{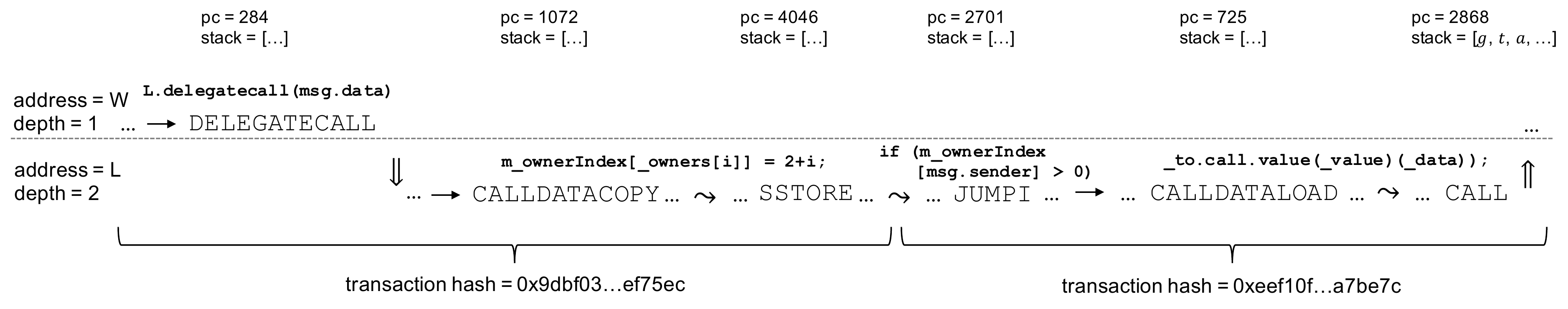}
  \caption{Execution example of an attack on an access control vulnerability. A data flow relation is depicted with $\leadsto$. The variables \textit{g}, \textit{t} and \textit{a} are as discussed in Figure \ref{fig:instructionflow}.}
  \label{fig:instructionflowaccesscontrol}
\end{figure*}
Reconsider the access control example described in Section~\ref{sec:reentrancy}. Figure~\ref{fig:instructionflowaccesscontrol} illustrates the relevant control flow, data flow and follows relations that occur during the execution of that example.
We note that the execution example is divided into two separate transactions. In the first transaction, the attacker sets itself as the owner, whereas in the second transaction the attacker transfers all the funds to itself.
Although in reality an attacker performs two separate transactions, in our methodology, the two transactions are represented as a single sequence of execution events. 
For both transactions, the execution starts with contract address $W$ eventually making a delegate call to contract address $L$, as part of the attacker calling the fallback function of contract $W$.
In the first transaction, we see that at a certain point contract $L$ copies data from the transaction using the \textcolor{teal}{\texttt{CALLDATACOPY}} instruction and stores it into storage via the \textcolor{teal}{\texttt{SSTORE}} instruction.
An interesting observation here is that state is shared across transactions through storage.
In the second transaction, the data that has previously been stored is now loaded onto the stack and used by a comparison. A comparison is ultimately reflected via the \textcolor{teal}{\texttt{JUMPI}} instruction. Finally, we see that the comparison follows a \textcolor{teal}{\texttt{CALLDATALOAD}} instruction whose data is used by a call \textcolor{teal}{\texttt{CALL}} instruction.
Given these observations, we are now able to create the following attack pattern in order to detect and thereby prevent the first Parity wallet hack:
\newline
\begin{rosettatable}
(opcode = DELEGATECALL) $\Rightarrow$ (opcode = CALLDATACOPY) $\leadsto$ 
(opcode = SSTORE) $\leadsto$ (opcode = JUMPI) where 
  (src.transaction.hash $\neq$ dst.transaction.hash) $\rightarrow$ 
((opcode = CALLDATALOAD) $\leadsto$ (opcode = CALL)) where 
  (dst.stack(2) > 0)
\end{rosettatable}

\noindent
\newline
The above attack pattern evaluates to true if the
following two conditions are met:
\begin{enumerate}
\item there is a transaction with a control flow relation between a \textcolor{teal}{\texttt{DELEGATECALL}} instruction and a \textcolor{teal}{\texttt{CALLDATACOPY}} instruction, where the data of the \textcolor{teal}{\texttt{CALLDATACOPY}} instruction flows into an \textcolor{teal}{\texttt{SSTORE}} instruction;
\item there is another transaction (i.e. \textcolor{violet}{\texttt{src}}.\textcolor{blue}{\texttt{transaction}}.\textcolor{blue}{\texttt{hash}} $\neq$ \textcolor{violet}{\texttt{dst}}.\textcolor{blue}{\texttt{transaction}}.\textcolor{blue}{\texttt{hash}}) where the data that has been previously stored via the \textcolor{teal}{\texttt{SSTORE}} instruction flows into a \textcolor{teal}{\texttt{JUMPI}} instruction and is followed by a \textcolor{teal}{\texttt{CALLDATALOAD}} instruction whose data flows into a \textcolor{teal}{\texttt{CALL}} instruction that sends out funds (i.e. \textcolor{violet}{\texttt{dst}}.\textcolor{blue}{\texttt{stack}}(2) > 0).
\end{enumerate}

\noindent
It is worth noting that the Parity wallet attack is a multi-transactional attack and that it is therefore significantly different from a reentrancy attack, that is solely based on a single transaction.
For more examples of attack patterns, please refer to Table~\ref{tbl:listofpatterns} in Appendix~\ref{sec:appendixa}.

\subsection{Decentralized Security Updates}

While our approach of using a DSL allows us to have a generic solution for
detecting attacks, it still leaves two open questions: 
\begin{enumerate}
    \item How do we distribute and enforce the same patterns across all the clients? 
    \item How do we decentralize the governance of patterns in order to prevent a single entity from deciding which patterns are added or removed?
\end{enumerate}

\begin{figure}
  \centering
  \includegraphics[scale=0.21]{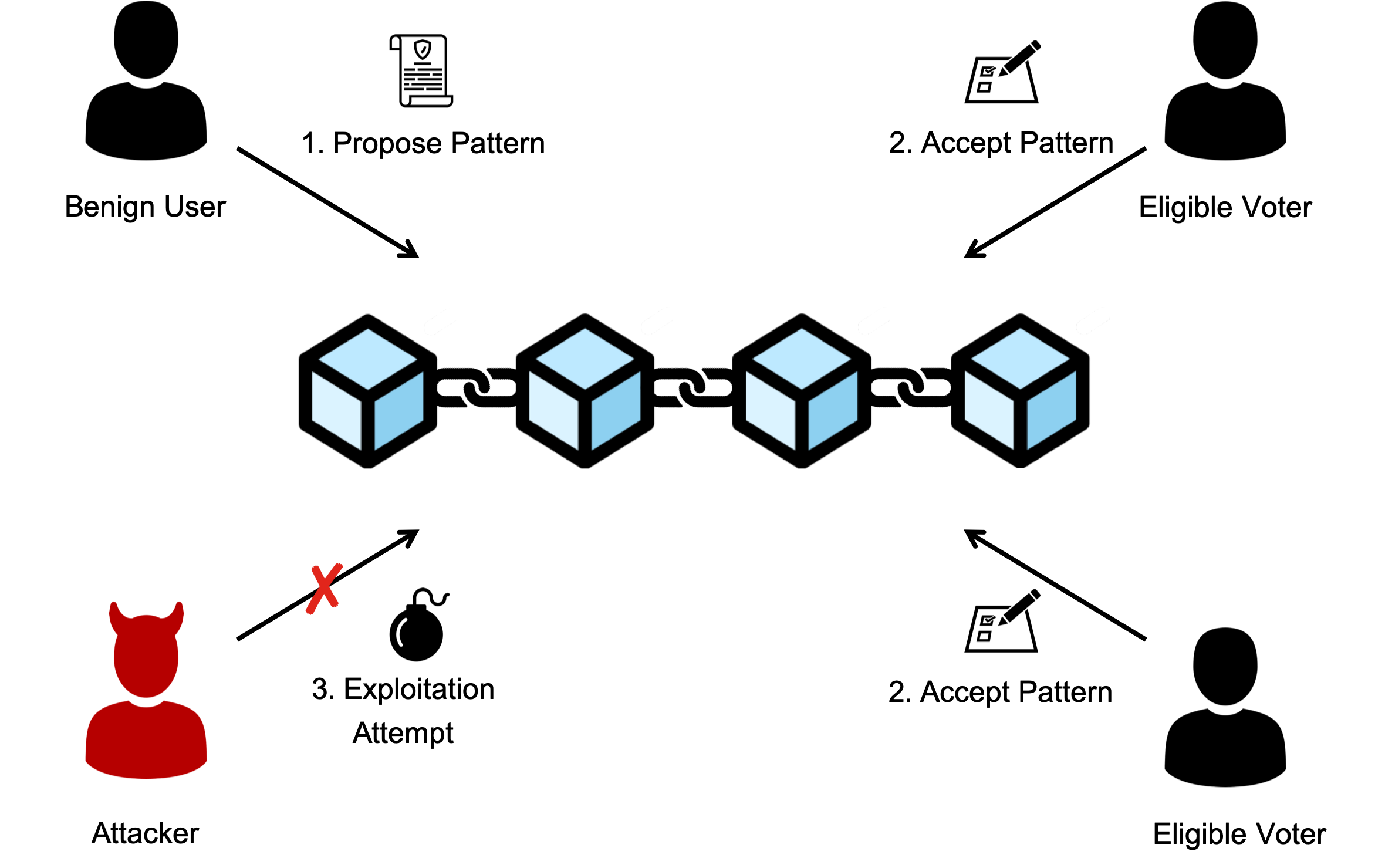}
  \caption{An illustrative example of \toolname's workflow: step 1) A
  benign user detects a vulnerability and proposes a pattern (written
  using our DSL) to the smart contract. Step 2) Eligible voters vote to
  either accept or reject the pattern. If the majority votes to accept the
  pattern, then all the clients are updated and the pattern is
  activated. Step 3) An attacker tries but fails to exploit a vulnerable
  smart contract due to the voted pattern matching the malicious
  transaction.}
  \label{fig:methodology}
\end{figure}

\noindent
The answer to these questions is to use a smart contract that is deployed on
the blockchain itself. This solves the problem of distributing and
enforcing that the same patterns are always used across all clients.
Specifically, patterns are stored inside the smart contract and the
blockchain protocol itself guarantees that every client knows about the
exact same state and therefore has access to exactly the same patterns.
The second problem of decentralizing the governance of patterns, is
solved by allowing the proposal and voting of patterns via the smart
contract as depicted in Figure~\ref{fig:methodology}. The contract maintains a
list of eligible voters that vote for either accepting or rejecting a
new pattern. If the majority has voted with ``yes'', i.e. to accept the pattern, then it is
added to the list of active patterns. In that case, every client is
automatically notified through the mechanism of smart contract events, and retrieves the updated list of patterns from
the blockchain. In other words, if a pattern is accepted by the voting
mechanism, it is updated across all the clients through the existing
consensus mechanism of the Ethereum blockchain. 
However, solving the second problem using a voting mechanism opens up a new problem concerning the requirements needed for governing the votes. 
In voting literature,
verifiability and privacy are typically seen as key requirements.
\emph{Verifiability} concerns linking the output to the input in a
verifiable way. \emph{Privacy} concerns whether a vote can be linked
back to a voter. In addition, we argue that the situation here is more
akin to boardroom voting than to general elections, because it should be
possible to hold voters \emph{accountable}. This means that privacy must
be maintained only until the election is over. Finally, the voting system must
not be favorable to any voters -- e.g., it should not confer an advantage to
voters that cast their vote late. This final property is called
\emph{fairness}. It is worth noting that fairness requires privacy during the voting
phase. This leads to the following three requirements:

\begin{enumerate}
\item \textbf{Verifiability:} The outcome of the vote must be verifiably
	related to the votes as cast by the voters;
\item \textbf{Accountability:} Voters can be held accountable for how
	they voted;
\item \textbf{Fairness:} No intermediate information must be leaked.
\end{enumerate}

\section{Implementation}

\begin{figure}
  \centering
  \includegraphics[scale=0.29]{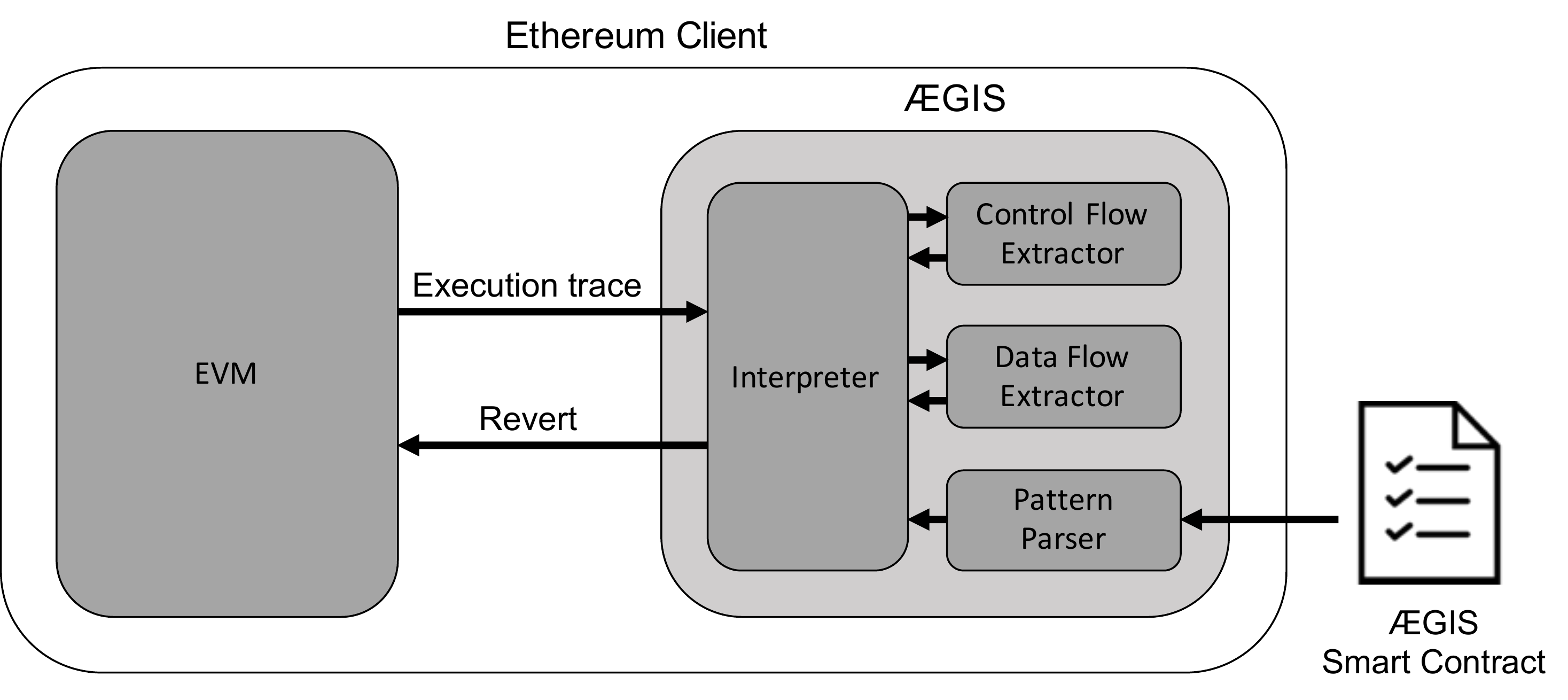}
  \caption{Architecture of \toolname{}. The dark gray boxes represent \toolname{}'s main components.}
  \label{fig:architecture}
\end{figure}

In this section, we provide the implementation details of our solution called \toolname{}. 
The code is publicly available\footnote{\url{https://github.com/christoftorres/Aegis}}.
Figure~\ref{fig:architecture}, provides an overview of the architecture of \toolname{} and highlights its main components. \toolname{} is implemented on top of Trinity\footnote{\url{https://trinity.ethereum.org/}}, an Ethereum client implemented in Python.

\subsection{Ethereum Client}

\noindent
\newline
\textbf{EVM.}
We modified the EVM of Trinity such that it keeps track of all the executed instructions and their states at runtime, in the form of an ordered list.
We refer to this list as the execution trace. 
Each record in this list contains the executed opcode, the value of the program counter, the depth of the call stack, the address of the contract that is being executed, and finally, all the values that were stored on the stack and in memory.
This list is passed to the interpreter component of \toolname{}.
\\
\newline
\textbf{Interpreter.} The interpreter loops through the list of executed instructions and passes the relevant instructions to the control flow and data flow extractor components. 
It is also responsible for signalling the EVM a revert in case the execution trace matches an attack pattern.
\\
\newline
\textbf{Control Flow Extractor.} The control flow extractor is responsible for inferring control flow information.
We do so by dynamically building a call tree from the instructions received by the interpreter.
A control flow relation is reported if there exists a path along the call tree, from the source instruction to the destination instruction defined in a given pattern.
Thus, control flow relations represent call dependencies between two instructions.
\\
\newline
\textbf{Data Flow Extractor.} The data flow extractor is responsible for collecting data flow information.
We track the flow of data between instructions by using dynamic taint analysis.
Taint is introduced whenever we come across a source instruction and checked whenever we come across a destination instruction. Source and destination instructions are defined by a given pattern.
Taint propagation follows the semantics of the EVM~\cite{wood2014ethereum} across stack, memory and storage.
We perform byte-level precision tainting. 
Taint that is stored across stack and memory is volatile, meaning that it is cleared across transactions. 
Taint that is stored across storage is persistent, meaning that it remains in storage across transactions.
This allows us to perform inter-transactional taint analysis.
A data flow relation is given if taint flows from a source instruction into a destination instruction.
\\
\newline
\textbf{Pattern Parser.} The pattern parser is responsible for extracting and parsing the patterns from the voting smart contract.
We implemented our pattern language using \text{textX} \footnote{\url{https://github.com/textX/textX}}, a Python framework providing a meta-language for building DSLs.

\subsection{\toolname{} Smart Contract}

The \toolname{} smart contract ensures proper curation of the list of active patterns. We implemented our smart contract in Solidity.
As previously mentioned, patterns are accepted or removed via a voting mechanism. The contract holds all proposed additions and removals of patterns and allows a vote on them within a set time window. The duration can be configured and updated by the contract owner. Proposals are open to the public and anyone can propose an addition to or removal from the list of patterns.

\begin{figure}[H]
  \centering
  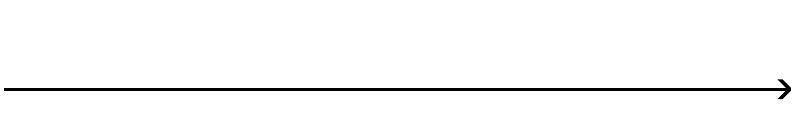
  \caption{Timeline of the two voting stages.}
  \label{fig:timeline}
\end{figure}

\noindent
\newline
\textbf{Fairness.}
Votes should remain secret until all eligible voters have had sufficient
opportunity to vote. Therefore, two time windows are employed.
The first window is for sending a commitment that includes a deposit.
The second window is for revealing a vote including the return of the committed deposits. The two windows are illustrated in Figure~\ref{fig:timeline}. In the figure, t\textsubscript{p} represents the point in time when a pattern is proposed and marks the start of the commit window. t\textsubscript{c} marks the end of the commit window and the start of the reveal window. Lastly, t\textsubscript{r} marks the end of the reveal window and the time when the pattern list is updated in case of a positive vote outcome. A commitment is a hash of the vote ID, the voter's vote and a nonce.
The vote ID is a hash of the proposed pattern and identifies the pattern that is being voted on.
The voter's vote is encoded as a boolean. 
The nonce ensures that commitments cannot be replayed. 
The smart contract records these commitments, which must be sent with the predefined deposit and within the predefined time window. During the commitment phase no one knows how anyone else has voted on a given pattern, and so cannot be swayed by the decisions of others. However, the process should ultimately be transparent to both voters and non-voters to foster trust in the system. As such, during the second window, the reveal window, all voters reveal how they have voted. They must reveal their vote in order to get their deposit back. No commits may be made once the reveal period has started.
\newline

\noindent
\textbf{Tallying.}
The voting ends either when more than 50\% (50\%+1 vote) of the total number of votes reaches either accept or reject, or when the time window for revealing expires with less than 50\% having been reached. In case the voting has ended but the reveal window has not yet passed, any remaining voters are still eligible to reveal their vote, such that their deposit can be returned. The reveal period is bounded so that patterns are accepted or rejected in a practical amount of time.
In the event of a successful vote, the pattern to which the vote pertains is added to or removed from the record held by the contract, according to the proposal. If a vote is unsuccessful, i.e. no majority voted for the proposal, the record of patterns is not changed.

\noindent
\newline
\textbf{Actors.}
There are three types of actors: the proposers that submit proposals to add or remove patterns, the voters that vote on proposals, and the admins that govern the list of eligible voters as well as the parameters of the smart contract (e.g. deposit, commit and reveal windows, etc.).
The \toolname{} smart contract allows every user on the blockchain to become a proposer by submitting a proposal.
Voters then vote on the proposals by first committing their vote and at a later stage revealing it.
Not every user is an eligible voter. Voters are only those users whose account address is stored in the list of eligible voters maintained by the smart contract.
Admins may update the list of eligible voters.
They oversee the proper curation of the smart contract and act as a governing body.
Admins are agreed upon off-chain and are represented by a multi-signature wallet. A multi-signature wallet is an account address which only performs actions if a group of users give their consent in form of a signature.

\noindent
\newline
\textbf{Data Structures.}
The smart contract consists of several functions and data structures that allow for the voting process to take place.
We make use of a number of \textit{modifiers}, which act as checks carried out before specific functions are executed. We use these to check that: 1) a voter is eligible, 2) a vote is in progress, 3) a reveal is in progress and 4) the associated vote has ended.
We use a struct to hold the details of each vote, these include the {\tt patternID}, the proposed {\tt pattern} and the {\tt startBlock}. 
These values enable us to record the details needed to check when a vote ends, check that the same pattern has not already been proposed, and count the number of votes. The struct is used in conjunction with a mapping, which maps a 32 bytes value to the details of each vote. The 32 bytes value represents the {\tt voteID} of each vote, created by hashing unique vote information.
A constructor is used to define, at contract launch, the value of the necessary deposit and the time windows during which voters can commit or reveal. The former is given in ether, while the latter are given in number of blocks. The deposit is used to ensure that those who committed a vote also reveal their vote. These values can be changed later using the contract's admin functions.

\noindent
\newline
\textbf{Functionality.}
The public functions for the voting process are: {\tt addProposal}, {\tt removeProposal}, {\tt commitToVote} and {\tt revealVote}. Both proposal functions first check if a vote with the same ID already exists, and if not create a new instance of voting details via the mapping.
Next, the {\tt commitToVote} function can be used inside the defined number of blocks to submit a unique hash of an eligible voter's vote. This function makes use of the {\tt canVote} modifier to protect access. The voter's commitment and vote hash are stored only if the correct deposit amount was sent to the function.
Once the vote stage has ended the reveal stage begins. During this window the {\tt revealVote} function, protected by the {\tt canVote} modifier, processes vote revelations and returns deposits. The function checks that the stored hash matches the hash calculated from the parameters passed to it, and if so, returns the voter's deposit and records the vote. Lastly, it calls an internal function which tallies the votes and adds or removes the pattern if either the for or against vote has reached over 50\%. In this way the vote is self tallying.
The patterns are ultimately stored in an array that can be iterated over to ensure each node has the full set.
Finally, the contract also has two admin functions: {\tt transferOwnership}, {\tt changeVotingWindows}. Both of these are protected by the {\tt isOwner} modifier. The former allows the current owning address to transfer control of the contract to a new address. The latter allows the commit and reveal windows to be changed as well as the amount required as a voting deposit.

\section{Evaluation}

In this section, we evaluate the effectiveness and correctness of \toolname, by conducting two experiments.
In the first experiment we compare the effectiveness of \toolname~to two state-of-the-art reentrancy detection tools: \textsc{ECFChecker}~\cite{grossman2017online} and \textsc{Sereum}~\cite{rodler2019sereum}.
In the second experiment we perform a large-scale analysis and measure the correctness as well as the performance of \toolname{} across the first 4.5 million blocks of the Ethereum blockchain.

\subsection{Comparison to Reentrancy Detection Tools}

\begin{table*}
\centering
    \begin{tabular}{l c c c c c c c c c c c c c c c c}
        & \rot{\text{CCRB}} 
        & \rot{\text{DAO}} 
        & \rot{\text{0x7484a1}} 
        & \rot{\text{proxyCC}} 
        & \rot{\text{DAC}}
        & \rot{\text{DSEthToken}}
        & \rot{\text{0x695d73}} 
        & \rot{\text{EZC}} 
        & \rot{\text{0x98D8A6}}
        & \rot{\text{WEI}}
        & \rot{\text{0xbD7CeC}} 
        & \rot{\text{0xF4ee93}} 
        & \rot{\text{Alarm}} 
        & \rot{\text{0x771500}} 
        & \rot{\text{KissBTC}} 
        & \rot{\text{LotteryGameLogic}} \\
        \toprule
        \text{\textsc{Sereum}} & \hspace{0.07cm}\FP\hspace{0.07cm} & \hspace{0.07cm}\TP\hspace{0.07cm} & \hspace{0.07cm}\FP\hspace{0.07cm} & \hspace{0.07cm}\FP\hspace{0.07cm} & \hspace{0.07cm}\FP\hspace{0.07cm} & \hspace{0.07cm}\TP\hspace{0.07cm} & \hspace{0.07cm}\FP\hspace{0.07cm} & \hspace{0.07cm}\FP\hspace{0.07cm} & \hspace{0.07cm}\FP\hspace{0.07cm} & \hspace{0.07cm}\FP\hspace{0.07cm} & \hspace{0.07cm}\FP\hspace{0.07cm} & \hspace{0.07cm}\FP\hspace{0.07cm} & \hspace{0.07cm}\FP\hspace{0.07cm} & \hspace{0.07cm}\FP\hspace{0.07cm} & \hspace{0.07cm}\FP\hspace{0.07cm} & \hspace{0.07cm}\FP\hspace{0.07cm} \\
        \midrule
        \toolname & \text{\TN} & \text{\TP} & \text{\TN} & \text{\TN} & \text{\TN} & \text{\TP} & \text{\TN} & \text{\TN} & \text{\TN} & \text{\TN} & \text{\TN} & \text{\TN} & \text{\TN} & \text{\TN} & \text{\TN} & \text{\TN} \\
     	\bottomrule
	\\
    \end{tabular}
    \caption{Comparison between \textsc{Sereum} and \toolname{} on the effectiveness of detecting reentrancy attacks.}
    \label{tbl:effectiveness}
\end{table*}

By analyzing transactions sent to contracts, Rodler et al.'s tool \textsc{Sereum} flagged 16 contracts as victims of reentrancy attacks.
However, after manual investigation the authors found that only 2 out of the 16 contracts have actually become victims to reentrancy attacks.
We decided to analyze these 16 contracts and see if we face the same challenges in classifying these contracts correctly.
We contacted the authors of \textsc{Sereum} and obtained the list of contract addresses.
Afterwards, we ran \toolname{} on all transactions related to the contract addresses, up to block number 4,500,000\footnote{This is the maximum block number analyzed by the authors of \textsc{Sereum}.}.
Table~\ref{tbl:effectiveness}~summarizes our results and provides a comparison to the results obtained by \textsc{Sereum}.
From Table~\ref{tbl:effectiveness}, we can observe that \toolname~successfully detects transactions related to the DAO contract and the DSEthToken contract, as reentrancy attacks.
Moreover, \toolname{} correctly flags the remaining 14 contracts as not vulnerable.
Hence, in contrast to \textsc{Sereum}, \toolname~produces no false positives on these 16 contracts.
After analyzing the false positives produced by \textsc{Sereum}, we conclude that \toolname{} does not produce the same false positives because first, \toolname{} does not use taint analysis in its pattern and therefore does not face issues of over-tainting, and secondly, it does not make use of dynamic write locks to detect reentrancy.

\begin{table}
\centering
    \begin{tabular}{l c c c c }
        Smart Contract & Reentrancy Type & \rot{\textsc{ECFChecker}} & \rot{\textsc{Sereum}} & \rot{\toolname{}} \\
        \toprule
        \multirow{2}{*}{\textit{VulnBankNoLock}} 		& \text{Same-Function} & \hspace{0.1cm}\TP\hspace{0.1cm} & \hspace{0.1cm}\TP\hspace{0.1cm} & \hspace{0.1cm}\TP\hspace{0.1cm} \\
         									& \text{Cross-Function} & \FN & \TP & \TP \\
\hline
        \multirow{2}{*}{\textit{VulnBankBuggyLock}}	& \text{Same-Function} & \TN & \FP & \TN \\
          									& \text{Cross-Function} & \FN & \TP & \TP \\
\hline
        \multirow{2}{*}{\textit{VulnBankSecureLock}}	& \text{Same-Function} & \TN & \FP & \TN \\
       	 							 		& \text{Cross-Function} & \TN & \FP & \TN \\
    	\bottomrule
	\\
    \end{tabular}
    \caption{Comparison between \textsc{ECFChecker}, \textsc{Sereum} and \toolname{} on the effectiveness of detecting same-function and cross-function reentrancy attacks with manually introduced locks.}
    \label{tbl:locks}
\end{table}

\subsubsection{Reentrancy with Locks}

Besides evaluating \textsc{Sereum} on the set of 16 real-world smart contracts, Rodler et al.~also compared \textsc{Sereum} to \textsc{ECFChecker}, using self-crafted smart contracts as a benchmark~\cite{rodler2019reentrancypatterns}.
The goal of this benchmark is to provide means to investigate the quality of reentrancy detection tools.
The benchmark consists of three functionally equivalent contracts, except that the first contract does not employ any locking mechanism to guard the reentry of functions (\textit{VulnBankNoLock}), the second contract employs partial implementation of a locking mechanism (\textit{VulnBankBuggyLock}), and the third contract employs a full implementation of a locking mechanism (\textit{VulnBankSecureLock}).
As a result, the first contract is vulnerable to same-function reentrancy as well as cross-function reentrancy. 
The second contract is vulnerable to cross-function reentrancy, but not to same-function reentrancy.
Finally, the third contract is safe regarding both types of reentrancy.
We deployed these three contracts on the Ethereum test network called Ropsten and ran the three contracts against \toolname{}.
Table~\ref{tbl:locks} contains our results and compares \toolname{} to \textsc{ECFChecker} and \textsc{Sereum}.
We can see that \textsc{ECFChecker} has difficulties in detecting cross-function reentrancy, whereas \textsc{Sereum} has difficulties in distinguishing between reentrancy and manually introduced locks.
This is probably due to the locking mechanism exhibiting exactly the same pattern as a reentrancy attack and \textsc{Sereum} being unable to differentiate between these two.
We found that \toolname~correctly classifies every contract as either vulnerable or not vulnerable in all the test cases.

\subsubsection{Unconditional Reentrancy}

Calls that send ether are usually protected by a check in the form of an \texttt{if}, \texttt{require}, or \texttt{assert}. 
Reentrancy attacks typically try to bypass these checks. 
However, it is possible to write a contract, which does not perform any check before sending ether. 
Rodler et al. present an example of such a vulnerability and name it \textit{unconditional reentrancy} (see Appendix~\ref{sec:appendixb}).
Moreover, they also find an example of such a contract deployed on the Ethereum blockchain\footnote{\url{https://etherscan.io/address/0xb7c5c5aa4d42967efe906e1b66cb8df9cebf04f7}}.
When \textsc{Sereum} was published, it was not able to detect this type of reentrancy since the authors assumed that every call that may lead to a reentrancy is guarded by a condition.
However, the authors claim to have fixed this issue by extending \textsc{Sereum} to tracking data flows from storage to the parameters of calls. 
We cannot verify this since the source code of \textsc{Sereum} is not publicly available. 
We run \toolname{} on both examples, the manually crafted example by Rodler et al. and the contract deployed on the Ethereum blockchain. 
\toolname{} correctly identifies the unconditional reentrancy contained in both examples without modifying the existing patterns. 
This is as expected, since in contrast to \textsc{Sereum}'s initial way to detect reentrancy, \toolname{}'s reentrancy patterns do not rely on the detection of conditions (i.e. \texttt{JUMPI}) to detect reentrancy.

\subsection{Large-Scale Blockchain Analysis}

In this experiment we analyse the first 4.5 million blocks of the Ethereum blockchain and compare our findings to those of Rodler et al.
We started by scanning the Ethereum blockchain for smart contracts that have been deployed until block 4,500,000.
We found 675,444 successfully deployed contracts. 
The deployment timestamps of the found contracts range from August 7, 2015 to November 6, 2017.
Next, we replayed the execution history of these 675,444 contracts.
As part of the scanning we found that only 12 contracts in our dataset have more than 10.000 transactions.
Therefore, to reduce the execution time, we decided to limit our analysis to the first 10.000 transactions of each contract. 
In addition, similar to Rodler et al., we tried our best to skip those transactions which were involved in denial-of-service attacks as they would result in high execution times\footnote{\url{https://tinyurl.com/rvlvues}}.

\begin{table}
\centering
\begin{tabular}{l r r }
Vulnerability & Contracts & Transactions \\
\toprule
Same-Function Reentrancy    & 7     & 822   \\
Cross-Function Reentrancy   & 5     & 695  \\
Delegated Reentrancy        & 0     & 0     \\
Create-Based Reentrancy     & 0     & 0     \\
Parity Wallet Hack 1        & 3     & 80    \\
Parity Wallet Hack 2        & 236   & 236   \\
\midrule
Total Unique                & 248   & 1118  \\
\bottomrule
\\
\end{tabular}
\caption{Number of vulnerable contracts detected by \toolname.}
\label{tbl:blockchainresults}
\end{table}

We ran \toolname{} on our set of 675,444 contracts using  a 6-core
Intel Core i7-8700 CPU @ 3.20GHz and 64 GB RAM. Our tool took on average 108 milliseconds to analyse a transaction, with a median of 24 milliseconds per transaction.
All in all, we re-executed 4,960,424 transactions with an average of 8 transactions per contract.
Table~\ref{tbl:blockchainresults} summarizes our results.
\toolname{} found a total of 1,118 malicious transactions and 248 unique contacts that have been exploited through either a reentrancy or an access control vulnerability.
More specifically, \toolname{} found that 7 contracts have become victim to same-function reentrancy, 5 contracts to cross-function reentrancy, 3 contracts to the first Parity wallet hack and 236 contracts to the second Parity wallet hack.
Similar to the results of Rodler et al., we did not find any contracts to have become victim to delegated reentrancy or create-based reentrancy.
We validated all our results by manually analyzing the source code (whenever it was publicly available) and/or the execution traces of the flagged contracts. 
Our validation did not reveal any false positives. 

\begin{table}[b]
\centering
\begin{tabular}{l c}
Contract Address & Block Range \\
\toprule
\small{\href{https://etherscan.io/address/0xd654bdd32fc99471455e86c2e7f7d7b6437e9179}{0xd654bdd32fc99471455e86c2e7f7d7b6437e9179}} & 1680024 - 1680238 \\
\small{\href{https://etherscan.io/address/0xbb9bc244d798123fde783fcc1c72d3bb8c189413}{0xbb9bc244d798123fde783fcc1c72d3bb8c189413}} & 1718497 - 2106624 \\
\cellcolor{gray!25}\small{\href{https://etherscan.io/address/0xf01fe1a15673a5209c94121c45e2121fe2903416}{0xf01fe1a15673a5209c94121c45e2121fe2903416}} & \cellcolor{gray!25}1743596 - 1743673 \\
\cellcolor{gray!25}\small{\href{https://etherscan.io/address/0x304a554a310c7e546dfe434669c62820b7d83490}{0x304a554a310c7e546dfe434669c62820b7d83490}} & \cellcolor{gray!25}1881284 - 1881284 \\
\cellcolor{gray!25}\small{\href{https://etherscan.io/address/0x59752433dbe28f5aa59b479958689d353b3dee08}{0x59752433dbe28f5aa59b479958689d353b3dee08}} & \cellcolor{gray!25}3160801 - 3160801 \\
\cellcolor{gray!25}\small{\href{https://etherscan.io/address/0xbf78025535c98f4c605fbe9eaf672999abf19dc1}{0xbf78025535c98f4c605fbe9eaf672999abf19dc1}} & \cellcolor{gray!25}3694969 - 3695510 \\
\cellcolor{gray!25}\small{\href{https://etherscan.io/address/0x26b8af052895080148dabbc1007b3045f023916e}{0x26b8af052895080148dabbc1007b3045f023916e}} & \cellcolor{gray!25}4108700 - 4108700 \\
\bottomrule
\\
\end{tabular}
\caption{Same-function reentrancy vulnerable contracts detected by \toolname{}. Contracts highlighted in gray have only been detected by \toolname{} and not by \textsc{Sereum}.}
\label{tbl:samefunctionreentrancyresults}
\end{table}

Table~\ref{tbl:samefunctionreentrancyresults} lists all the contract addresses that \toolname{} detected to have become victim of a same-function reentrancy attack. 
The block range defines the block heights where \toolname{} detected the malicious transactions. 
The first and second contract addresses contained in Table~\ref{tbl:samefunctionreentrancyresults} are the same as reported by \textsc{Sereum}, and belong to the DSEthToken and DAO contract, respectively.
The rows highlighted in gray mark 5 contracts that have been flagged by \toolname{} but not by \textsc{Sereum}.
After investigating the transactions of these 5 contracts, we find that the contract addresses \texttt{0x26b8af052895080148dabbc1007b3045f023916e} and \texttt{0xbf7802
5535c98f4c605fbe9eaf672999abf19dc1} became victim to same-function reentrancy, but seem to be contracts that have been deployed with the purpose of studying the DAO hack. 
However, the three other contract addresses seem to be true victims of reentrancy attacks.

\section{Discussion}

In this section we discuss alternatives to determine eligible voters, highlight some of the current limitations as well as future research directions for this work.

\subsection{Determining Eligible Voters}

The introduction of new patterns in \toolname{} depends on achieving consensus
in a predetermined group of voters.
Although it may intuitively make sense to let miners vote,
they are not necessarily a good fit. 
Their interests may differ from those of smart contract users. 
For example, depending on a pattern's complexity, it might introduce an overhead in terms of execution time.
Miners are then incentivized to prefer simpler patterns that are
evaluated quicker, while smart contract users would prefer more secure patterns.

Alternatively, a group of trusted security experts could act as eligible
voters\footnote{Somewhat similar to how CVEs are handled.}.
Security experts are (by definition) able to properly evaluate patterns and have the
interest in doing so. The voting contract is then controlled by a group
of trusted experts who are decided upon off-chain by a group of admins. 
For transparency, the identity of admins and experts would be exposed to the public by mapping every identity to an Ethereum account. 
Changes to the list of voters, the deposit, or the commit and reveal windows are then visible to anyone via the blockchain. 
Through this setup,
security experts would be able to organise themselves with the voter list
being comprised of a curated group of knowledgeable people. Such groups
already exist in reality, for example, the members
of the Smart Contract Weakness Classification registry
(SWC)\footnote{https://smartcontractsecurity.github.io/SWC-registry/}, and would be a good fit for our system.
Moreover, misbehaving or unresponsive experts could be easily removed by the group of admins.
Although this approach allows for scalability and control, it has the disadvantage of introducing managing third-parties.
That runs counter to the decentralised concept of Ethereum.

Alternatively, there is also an option to select voters, while preserving the decentralised concept of Ethereum.
This is to remove the role of admins altogether, and instead
follow a self-organizing strategy, similar to Proof-of-Stake.
In this case, everyone is allowed to become a voter through the
purchase of (not prohibitively priced) voting power. This could
be achieved by depositing a fixed amount of ether into the
voting smart contract as a form of collateral. 

\subsection{Adoption and Participation Incentives}

The deployment of \toolname{} would require a modification of the Ethereum consensus protocol, which would require existing Ether-eum clients to be updated.
This could be easily achieved though a major release by including this one-time modification as part of a scheduled hardfork.
Another issue concerns the incentives to propose and vote on patterns. While prestige or a feeling of contributing
to the security of Ethereum may be sufficient for some, more
incentives may be needed to ensure that the protective capabilities
of \toolname{} are used to the full extent. A monetary
incentive could address this. That is, \toolname{} could be
extended with automatically paid rewards. In other words,
\toolname{} could be extended to enable bug bounties~\cite{breidenbach2018enter}.
\toolname{}'s smart contract could be modified such that, owners of smart contracts can register their contract address by sending a transaction to \toolname{}'s voting smart contract and deposit a bounty in the form of ether.
Then, proposers of patterns would be rewarded automatically with the bounty by \toolname{}'s voting smart contract, if their proposed pattern is accepted by the group of voters.
Moreover, owners could simply replenish the bounty for their contract by making new deposits to \toolname{}'s smart contract.

\subsection{Limitations and Future Work}

A current limitation of our tool is that proposed attack patterns are submitted in plain text to the smart contract.
Potential attackers can view the patterns and use them to find vulnerable smart contracts.
To mitigate this, we propose to make use of encryption such that only the voters would be able to view the patterns.
However, this would break the current capability of the smart contract being self-tallying. 
Designing an encrypted and practical self-tallying solution is left for future work.
Finally,
we intend to make use of parallel execution inside the extractors and the checking of patterns in order to improve the time required to analyse transactions.

 \section{Conclusion}

Although academia proposed a number of tools to detect vulnerabilities in smart contracts, they all fail to protect already deployed vulnerable smart contracts.
One of the proposed solutions is to modify the Ethereum clients in order to detect and revert transactions that try to exploit vulnerable smart contracts.
However, these solutions require all the Ethereum clients to be modified every time a new type of vulnerability is discovered.
In this work, we introduced \toolname{}, a system that detects and reverts attacks via attack patterns.
These patterns describe malicious control and data flows through the use of a novel domain-specific language.
In addition, we presented a novel mechanism for security updates that allows these attack patterns to be updated quickly and transparently via the blockchain, by using a smart contract as means of storing them.
Finally, we compared \toolname{} to two current state-of-the-art online reentrancy detection tools.
Our results show that
\toolname{} not only detects more attacks, but also has no false positives as compared to current state-of-the-art.

%
\begin{acks}
We would like to thank the \textsc{Sereum} authors, especially Michael Rodler, for sharing their data with us. 
We would also like to thank the reviewers for their valuable comments as well as Daniel Xiapu Luo for his valuable help. 
The experiments presented in this paper were carried out
using the HPC facilities of the University of Luxembourg~\cite{VBCGHPCS14}
{\small -- see \url{https://hpc.uni.lu}}.
This work is partly supported by the Luxembourg National Research Fund (FNR) under grant 13192291.
\end{acks}

%
\bibliographystyle{ACM-Reference-Format}
\bibliography{bibliography}

%
\appendix

\section{Complete List of \toolname's Attack Patterns}
\label{sec:appendixa}

Table~\ref{tbl:listofpatterns} provides a complete list of vulnerabilities as well as their respective attack patterns that \toolname{} is currently capable to detect.

\begin{table*}
\centering
\begin{tabular}{c l} 
Vulnerability & \multicolumn{1}{c}{Attack Pattern} \\
\toprule
\makecell{Same-Function \\ Reentrancy} & 
{\begin{lstlisting}[language=rosetta,numbers=none]
(opcode = CALL) $\Rightarrow$ (opcode = CALL) where (src.stack(1) = dst.stack(1)) $\wedge$ 
  (src.address = dst.address) $\wedge$ (src.pc = dst.pc) $\rightarrow$ 
(opcode = SSTORE) $\rightarrow$ (opcode = SSTORE) where (src.stack(0) = dst.stack(0)) $\wedge$ 
  (src.address = dst.address) $\wedge$ (src.depth > dst.depth)
\end{lstlisting}} \\
\midrule
\makecell{Cross-Function \\ Reentrancy} & 
{\begin{lstlisting}[language=rosetta,numbers=none]
(opcode = CALL) $\Rightarrow$ (opcode = CALL) where 
  (src.stack(1) = dst.stack(1)) $\wedge$ (src.address = dst.address) $\wedge$
  (src.memory(src.stack(3), src.stack(4)) $\neq$ dst.memory(dst.stack(3), dst.stack(4))) $\rightarrow$ 
(opcode = SSTORE) $\rightarrow$ (opcode = SSTORE) where (src.stack(0) = dst.stack(0)) $\wedge$ 
  (src.address = dst.address) $\wedge$ (src.depth > dst.depth)
\end{lstlisting}} \\
\midrule
\multirow{5}{*}{\makecell{Delegated \\ Reentrancy}} & 
{\begin{lstlisting}[language=rosetta,numbers=none]
(opcode = DELEGATECALL) $\Rightarrow$ (opcode = DELEGATECALL) where (src.stack(1) = dst.stack(1)) $\wedge$ 
  (src.address = dst.address) $\wedge$ (src.pc = dst.pc) $\rightarrow$ 
(opcode = SSTORE) $\rightarrow$ (opcode = SSTORE) where (src.stack(0) = dst.stack(0)) $\wedge$ 
  (src.address = dst.address) $\wedge$ (src.depth > dst.depth)
\end{lstlisting}} \\ \cmidrule{2-2}
& {\begin{lstlisting}[language=rosetta,numbers=none]
(opcode = CALLCODE) $\Rightarrow$ (opcode = CALLCODE) where (src.stack(1) = dst.stack(1)) $\wedge$ 
  (src.address = dst.address) $\wedge$ (src.pc = dst.pc) $\rightarrow$ 
(opcode = SSTORE) $\rightarrow$ (opcode = SSTORE) where (src.stack(0) = dst.stack(0)) $\wedge$ 
  (src.address = dst.address) $\wedge$ (src.depth > dst.depth)
\end{lstlisting}} \\
\midrule
\makecell{Create-Based \\ Reentrancy} & 
{\begin{lstlisting}[language=rosetta,numbers=none]
(opcode = CREATE) $\Rightarrow$ (opcode = CREATE) where (src.stack(1) = dst.stack(1)) $\wedge$ 
  (src.address = dst.address) $\wedge$ (src.pc = dst.pc) $\rightarrow$ 
(opcode = SSTORE) $\rightarrow$ (opcode = SSTORE) where (src.stack(0) = dst.stack(0)) $\wedge$ 
  (src.address = dst.address) $\wedge$ (src.depth > dst.depth)
\end{lstlisting}} \\
\midrule
\makecell{Parity Wallet \\ Hack 1} & 
{\begin{lstlisting}[language=rosetta,numbers=none]
(opcode = DELEGATECALL) $\Rightarrow$ (opcode = CALLDATACOPY) $\leadsto$ (opcode = SSTORE) $\leadsto$ (opcode = JUMPI) where 
  (src.transaction.hash $\neq$ dst.transaction.hash) $\rightarrow$ 
((opcode = CALLDATALOAD) $\leadsto$ (opcode = CALL)) where 
  (dst.stack(2) > 0)
\end{lstlisting}} \\
\midrule
\makecell{Parity Wallet \\ Hack 2} & 
{\begin{lstlisting}[language=rosetta,numbers=none]
(opcode = CALLDATACOPY) $\leadsto$ (opcode = SSTORE) $\leadsto$ (opcode = JUMPI) where 
  (src.transaction.hash $\neq$ dst.transaction.hash) $\rightarrow$ 
((opcode = CALLDATALOAD) $\leadsto$ (opcode = SELFDESTRUCT))
\end{lstlisting}} \\
\midrule
\makecell{Integer Overflow \\ (Addition)} & 
{\begin{lstlisting}[language=rosetta,numbers=none]
(opcode = CALLDATALOAD) $\leadsto$ (opcode = ADD) where
  ((dst.stack(0) + dst.stack(1)) $\neq$ dst.stack.result) $\leadsto$ (opcode = CALL)
\end{lstlisting}} \\
\midrule
\makecell{Integer Overflow \\ (Multiplication)} & 
{\begin{lstlisting}[language=rosetta,numbers=none]
(opcode = CALLDATALOAD) $\leadsto$ (opcode = MUL) where
  ((dst.stack(0) * dst.stack(1)) $\neq$ dst.stack.result) $\leadsto$ (opcode = CALL)
\end{lstlisting}} \\
\midrule
\makecell{Integer Underflow} & 
{\begin{lstlisting}[language=rosetta,numbers=none]
(opcode = CALLDATALOAD) $\leadsto$ (opcode = SUB) where
  ((dst.stack(0) - dst.stack(1)) $\neq$ dst.stack.result) $\leadsto$ (opcode = CALL)
\end{lstlisting}} \\
\midrule
\makecell{Timestamp \\ Dependence} & 
{\begin{lstlisting}[language=rosetta,numbers=none]
(opcode = TIMESTAMP) $\leadsto$ (opcode = JUMPI) $\rightarrow$ (opcode = CALL) where (dst.stack(2) > 0)
\end{lstlisting}} \\
\midrule
\makecell{Transaction \\ Order Dependency} & 
{\begin{lstlisting}[language=rosetta,numbers=none]
(opcode = SSTORE) $\leadsto$ (opcode = SLOAD) where 
  (src.block.number = dst.block.number) $\wedge$ (src.transaction.from $\neq$ dst.transaction.from) \end{lstlisting}} \\
\bottomrule
\\
\end{tabular}
\caption{List of vulnerabilities and their respective attack patterns.}
\label{tbl:listofpatterns}
\end{table*}

\section{Unconditional Reentrancy Example}
\label{sec:appendixb}

Figure~\ref{fig:unconditionalreentrancy} shows an example of a smart contract with an unconditional reentrancy. In this example an attacker first deposits a small amount of ether and then uses a reentrancy attack in order to drain all the ether that every single user has deposited. 

\begin{figure}[H]
\begin{centering}
\begin{lstlisting}[language=Solidity]
contract VulnBank {
  mapping (address => uint) public userBalances;
  
  function deposit() public payable {
    userBalances[msg.sender] += msg.value;
  }
  
  function withdrawAll() public {
    uint amountToWithdraw = userBalances[msg.sender];
    msg.sender.call.value(amountToWithdraw)("");
    userBalances[msg.sender] = 0;
  }
}
\end{lstlisting}
\caption{Example of a contract that is vulnerable to unconditional reentrancy~\cite{rodler2019reentrancypatterns}.
}
\label{fig:unconditionalreentrancy}
\end{centering}
\end{figure}

\end{document}